\documentclass[12pt]{amsart}
\usepackage{amssymb}
\def\*{\,\cdot\,}

\def\g{\mathbf{g}}
\def\s{\mathbf{s}}
\def\Uh{\mathbf{u}}
\def\Vh{\mathbf{v}}
\def\x{\mathbf{x}}
\def\y{\mathbf{y}}
\def\p{\mathbf{p}}
\def\q{\mathbf{q}}

\def\a{\mathbf{a}}
\def\N{\mathbf{N}}

\def\nsites{F}
\def\EXP{\mbox{{\large\bf e}}}
\def\ds{\displaystyle}
\def\tr{\mbox{tr}}
\newtheorem{prop}{Proposition}

\begin{document}

\title[] {Relativistic Toda chain\thanks{This work was supported by RFBR
grant No.
98-01-00070}}

\author{G. Pronko \and S. Sergeev}

\thanks{This work was supported by RFBR grant No. 98-01-00070}%

\address{Institute for High Energy Physics, Protvino, Russia.}
\email{pronko@mx.ihep.su}
\address{Joint Institute for Nuclear Research,
Dubna, Russia.}
\email{sergeev@thsun1.jinr.ru}

\date{September 11, 2000}



\begin{abstract}
Investigated is the relativistic periodic Toda chain, to each
site of which the ultra-local Weyl algebra is associated. Weyl's
$q$ we are considering here is restricted to be inside the unit
circle. Quantum Lax operators of the model are intertwined by six
vertex $R$-matrix. Both independent Baxter's $Q$-operators are
constructed explicitly as seria over local Weyl generators. The
operator-valued wronskian of $Q$-s is also calculated.
\end{abstract}

\maketitle

\section{Introduction}

Long ago, in his famous papers \cite {Baxter} R.Baxter has
introduced the object, which is known now as $Q$-operator.This operator
does satisfy the so-called Baxter, or $T-Q$, equation and besides has many
interesting properties. Recently
$Q$-operator was intensively discussed in the series of papers
\cite {BLZ} in the connection with continuous quantum field
theory. In \cite {Sklyanin1,Sklyanin3} it was pointed out the
relation of $Q$-operator with quantum B\"aklund transformations.
In \cite{GP} was discovered the relation of $Q$-operator with Bloch
solutions of quantum linear problem.

$Q$-operator was used initially for the solution of the
eigenvalue problem of $XYZ$-spin chain, where usual Bethe ansatz
fails. The reason is that $T-Q$ equation, together with an
appropriate boundary conditions, provides an one-dimensional
multiparameter spectral problem which allows one to determine the
spectra both of the auxiliary transfer matrix $T$ and of the
operator $Q$. In the case of the quantum mechanical integrable
chains, e. g. the periodic Toda chain, the appropriate solution
of the Baxter equation plays the prominent role in the functional
Bethe ansatz and the quantum separation of variables.

In quite recent papers there obtained explicit constructions of
$Q$-operators for several models, like the isotropic Heisenberg
spin chain, \cite {Pronko}, the periodic Toda chain and other
models with the rational $R$-matrix, \cite{GP}. In these papers
$Q$-operator ({\it operator, but not the solution of the Baxter
equation}) was obtained as the trace of monodromies of the appropriate
local operators. It is well known, that with free boundary conditions
for $Q$, $T-Q$ equation provides an one-parametric family of
solutions, so that one may extract two independent solutions with
nonzero discrete wronskian, see \cite {BLZ} and \cite {PS}. In
the papers \cite{Pronko,GP} both independent $Q$ operators were
obtained for the models considered.

In this paper we investigate the exactly integrable model known as
``the relativistic Toda chain'',
\cite{Ruijsenaars,Kuznetsov,Kundu}. Local $L$ operator for the
model is constructed with the help of the Weyl algebra
generators, commuting on $q$, and we deal with the case
$|q|\;<\;1$. In this paper we do not consider the Jacoby partners
to the Weyl algebra, dealing thus with the compact $q$ --
dilogarithms. Quantum space of our model is a formal module of an
enveloping of the tensor product of several copies of Weyl
algebras. The only thing we suggest for the Weyl generators is
their invertibility and a $q$-equidistant spectrum for one of
them. Both independent operators $Q_+$ and $Q_-$ and their
wronskian are calculated locally as the operators acting in the
ultra-local Weyl algebra. {\it Actually all our results are to be
understood as the well defined series expansions for functions
from the enveloping mentioned.}

\section{The model and the results}

This section consists of two parts. We formulate the model at
first, actually just defining the transfer matrix, and then we
give the final formulae for $Q_\pm$ operators and their $q$ --
Wronskian. All the sections beyond the introduction are the
QUISM-type derivation of these results.

\subsection{Problem}

First of all let us define the relativistic Toda chain
$L$-operator as
\begin{equation}\label{L-operator}
\ds
L_f(x)\;=\;\left(\begin{array}{ccc}
\ds x\Uh_f\,-\,(x\,\Uh_f)^{-1} &,&\ds \Vh_f\\
&&\\
\ds q^{-1/2}\,\lambda\,\Vh_f^{-1} &,& 0\end{array}\right)\;,
\end{equation}
where $\{\Uh_f,\Vh_f\}$ form the ``half-integer'' ultra-local
Weyl algebra:
\begin{equation}\label{1/2weil}
\ds\Uh_f^{}\*\Vh_f^{}\;=\;q^{1/2}\;\Vh_f^{}\*\Uh_f^{}\;,
\end{equation}
and elements with different $f$-s commute. As usual, the whole
quantum space is the tensor product of some copies of Weyl
modules, and $f$ marks the ``number'' of given Weyl algebra in
this tensor product. Recall, we will always imply $|q|\;<\;1$.

The correspondence between the relativistic Toda chain and usual
Toda chain may be established, for example, in the following
parameterization
\begin{equation}
\ds q =\EXP^{-\,i\,\epsilon}\;,\;\; \lambda = -\epsilon^2\;,\;\;
x = \EXP^{\epsilon\,\theta/2}\;,\;\; \Uh_f =
\EXP^{-\,\epsilon\,\p_f/2}\;,\;\; \Vh_f = \epsilon\,\EXP^{\q_f}\;,
\end{equation}
where
\begin{equation}
\ds
[\,\p\;,\;\q\,]\;=\;i\;,
\end{equation}
in the limit
\begin{equation}
\ds
\lim_{\epsilon=0}\;{1\over \epsilon}\;L_f(x)\;=\;
\left(\begin{array}{ccc}
\ds \theta-\p_f &,& \ds\EXP^{\q_f}\\
&&\\
\ds-\,\EXP^{-\q_f} &,& \ds 0
\end{array}\right)\;.
\end{equation}
The right hand side of this relation is known as the Toda
$L$ operator.

In the $L$-operator as well as in all
other objects the spectral parameter
$x$ will always couple with $\Uh_f$.
So we introduce the useful notation:
\begin{equation}\label{xsf}
\ds x^2\;\Uh_f^2\;\stackrel{def}{=}\;q^{\s_f}\;,
\end{equation}
so that for any formal function $g(\s_f)$
\begin{equation}
\ds g(\s_f)\*\Vh_f^n\;=\;\Vh_f^n\*g(\s_f+n)\;\;\;\;\forall\;n\;.
\end{equation}
We define the transfer matrix for the chain with $\nsites$ sites,
$f=1,...,\nsites$, as
\begin{equation}\label{T-matrix}
\ds
T(x^2)\;=\;\left((-\,x)^\nsites\,\prod_f\;\Uh_f\right)\*
\tr\;\left(L_1(x)\*L_2(x)\;\cdots\;L_{\nsites}(x)\right)\;.
\end{equation}
$T(x^2)$ becomes a polynomial of $x^2$ with commutative coefficients:
\begin{equation}\label{T-decomposed}
\ds
T(x^2)\;=\;\sum_{j=0}^{\nsites}\;(-\,x^2)^{\nsites-j}\; t_j^{}\;.
\end{equation}
Here it is implied $t_\nsites=1$ and
\begin{equation}\label{t0}
\ds t_0\;=\;\prod_f\,\Uh_f^2\;.
\end{equation}
Note that apart from the trivial $t_\nsites\;=\;1$ all other $\nsites$
coefficients are independent.
For given set $\{t_j\}$ one can define another set $\{\overline{t_j}\}$ by
\begin{equation}
\ds\overline{t_j^{}}\;=\;t_0^{-1}\;t_{\nsites-j}^{}\;.
\end{equation}
This means simply
\begin{equation}
\ds
T(x^2)\;=\;(-\,x^2)^{\nsites}\,t_0\;\overline{T}(x^{-2})\;.
\end{equation}

We shall fix now the coefficients in Baxter's equation:
\begin{equation}\label{Baxter}
\ds
T(x^2)\;Q(x^2)\;=\;
\left((-\,\lambda\,x^2)^\nsites\;t_0\right)\;Q(q\,x^2)\;+\;
Q(q^{-1}x^2)\;,
\end{equation}
where $t_0$ is given by (\ref{t0}).
In what follows we shall see that with this normalization of the
coefficients in(\ref{Baxter}) the Baxter equation has a solution
entire on $x^2$. We shall call this solution
\begin{equation}\label{Q+J}
\ds Q_+(x^2)\;=\;J(x^2,\lambda,\{t\})\;.
\end{equation}
\begin{prop}
Entire on $x^2$ solution of (\ref{Baxter}) as a series on
$\lambda^\nsites$
is
\begin{equation}\label{Jseries}
\ds J(x^2,\lambda,\{t\})\;=\;
\left(\prod_{k=1}^\infty\;T(q^k\,x^2)\right)\*\left(\sum_{k=0}^\infty\;(-\
,\lambda^\nsites)^k\;c_k(x^2)\right)\;,
\end{equation}
where $c_{-1}\;\equiv\;0$, $c_{0}\;\equiv\;1$ and recursively
\begin{equation}\label{recursion}
\ds
c_{k}(x^2)\;=\;\sum_{j=1}^\infty\;{(q^j\,x^2)^\nsites\,c_{k-1}(q^{1+j}\,x^
2)\over
T(q^j\,x^2)\,T(q^{1+j}\,x^2)}\;.
\end{equation}
\end{prop}
Note, $J(x^2,\lambda,\{t\})$ is the entire function on all its
arguments. The proof of this proposition is rather simple
exercise.

The other solution $Q_-(x^2)$ must contain a cut with respect to $x$, and
up to this cut
we guess $Q_-(x^2)$ to be entire on $x^{-2}$. More
exactly, with the $\s_f$ -- notation introduced by eq. (\ref{xsf}),
\begin{equation}\label{Q-J}
\ds
Q_-(x^2)\;=\;\lambda^{-\sum_f\,\s_f}\*
J(x^{-2},\lambda,\{\overline{t}\})\;.
\end{equation}
The last definition we need is
the $q$-Wronskian of these two solutions:
\begin{equation}\label{W-definition}
\ds
W(x^2)\;\stackrel{def}{=}\;Q_+(q^{-1}x^2)\;Q_-(x^2)\;-\;Q_+(x^2)\;Q_-(q^{-
1}x^2)\;.
\end{equation}

\subsection{Solution}

In this paper we'll give explicit expressions for both functions
$Q_\pm$. The natural question arises: we've got yet the form
(\ref{Jseries}) and (\ref{recursion}), what one may do else. The
aim of this paper is to investigate the the relativistic Toda
chain by QUISM method, to construct local operators $M_f(x^2)$
such that a trace of their monodromy gives $Q_\pm(x^2)$,  to
prove the commutativity of the transfer matrices and $Q_\pm$ and
to calculate the Wronskian. Note that in QUISM approach we'll
construct $Q_\pm(x^2)$ not as functions of $\{t\}$, but as
functions of local $\Uh_f,\Vh_f$. This is in some sense a
factorization, the simplest analogue of this is well known
$q$-exponential formula
\begin{equation}\label{chou}
\ds (\x+\y;q)_\infty\;=\;(\x;q)_\infty\*(\y;q)_\infty;,\;\;\;
\x\;\y\;=\;q\;\y\;\x\;,
\end{equation}
where conventionally
\begin{equation}
\ds
(x;q)_n\;\stackrel{def}{=}\;\prod_{k=0}^{n-1}\;(1\,-\,q^k\,x)\;,\;\;\;
(x;q)_\infty\;\stackrel{def}{=}\;\prod_{n=0}^{\infty}\;(1\,-\,q^k\,x)\;,
\end{equation}
and as the series expansions
\begin{equation}
(x;q)_\infty\;=\;\sum_{n=0}^\infty\;q^{n\,(n-1)/2}\;{(-x)^n\over
(q;q)_n}\;,\;\;\;
(x;q)_\infty^{-1}\;=\;
\sum_{n=0}^\infty\;{x^n\over (q;q)_n}\;,
\end{equation}
The right hand side of eq. (\ref{chou}) we call the local
form of its ``global'' left hand side.

Now we describe the local form of all solutions. First of all we introduce
an
universal function
\begin{equation}\label{g-function}
\ds
\g_{\alpha,\beta}(n,m)\;\stackrel{def}{=}\;
q^{n\,m}\;\alpha^n\;\beta^m\;{(q^{1+n};q)_\infty\;
(q^{1+m};q)_\infty\over(q;q)_\infty}\;,
\end{equation}
where $\alpha$ and $\beta$ are complex numbers,
and elements $q^n$ and $q^m$ commute.
\begin{prop}
Operator $Q_{+}(x^2)$,  defined by eqs.
(\ref{Q+J},\ref{Jseries},\ref{recursion}), in the local form is
\begin{equation}\label{Q+}
\ds
Q_+(x^2)\;=\;
\sum_{\{n_f\geq0\}}\;
\left(\prod_{f}\;\g_{1,\lambda}(n_f+\s_f,n_f)\right)\*\left(\prod_f\;\left
(\Uh\Vh\right)_f^{n_{f+1}-n_f}\right)\;,
\end{equation}
and operator $Q_{-}(x^2)$,  defined by eqs.
(\ref{Q-J},\ref{Jseries},\ref{recursion}), in the local form is
\begin{equation}\label{Q-}
\ds
Q_-(x^2)\;=\;\sum_{\{n_f\geq 0\}}\;
\left(\prod_{f}\;\g_{1,\lambda}
(n_f,n_f-\s_f)\right)\*\left(\prod_f\;
\left(\Uh\Vh\right)_f^{n_f-n_{f-1}}\right)\;.
\end{equation}
Their Wronskian, defined by eq. (\ref{W-definition}), is
\begin{equation}\label{W}
\ds W(x^2) = \left(\prod_f(q^{\s_f};q)_\infty
(q^{1-\s_f};q)_\infty \lambda^{-\s_f}\right)\cdot
\left(\prod_{f}\left({\lambda (\Uh\Vh)_f\over
(\Uh\Vh)_{f+1}};q\right)_\infty\right).
\end{equation}
\end{prop}
This paper is actually the proof of the second proposition.

\section{Intertwiners}

\subsection{Integrability}

First of all, from what follows the integrability of the the
relativistic Toda chain, namely the commutativity of the transfer
matrices (\ref{T-matrix}). The origin of it is the famous
six-vertex $R$ matrix. The following relation holds:
\begin{equation}\label{RLL}
\ds
R_{1,2}(x/y)\* L_{1,f}(x)\* L_{2,f}(y)\;=\;
L_{2,f}(y)\* L_{1,f}(x)\* R_{1,2}(x/y)\;,
\end{equation}
where $\ds L_{1,f}(x)\;=\;L_f(x)\otimes 1$,
$\ds L_{2,f}(y)\;=\;1\otimes L_f(y)$ etc., the cross
product implies the tensor product of the $2\times 2$ matrices,
and the six-vertex $R$ matrix has the form
\begin{equation}
\ds
R(x)=\left(\begin{array}{cccc}
1-x^{-2}q & 0 & 0 & 0\\
&&&\\
0 & q^{1/2}(1-x^{-2}) & x^{-1}(1-q) & 0\\
&&&\\
0 & x^{-1}(1-q) & q^{1/2}(1-x^{-2}) & 0 \\
&&&\\
0 & 0 & 0 & 1-x^{-2}q
\end{array}\right).
\end{equation}
The Yang-Baxter relation (\ref{RLL}) provides the commutativity
of the traces of the monodromies for $L_{1,f}(x)$
and $L_{2,y}(x)$, and as the extra multiplier
in our definition of the transfer matrix (\ref{T-matrix})
is the total shift operator, our modified transfer matrices
(\ref{T-matrix}) also form the commutative family:
\begin{equation}
\ds
T(x^2)\*T(y^2)\;=\;T(y^2)\*T(x^2)\;.
\end{equation}

The appearance of the six-vertex $R$ matrix is the criterion of
the existence of Baxter's ``$TQ=Q'+Q''$'' relation for our
transfer matrix. Let $M_{h,f}(x^2)$ be an operator, acting in
tensor product of $f$-th quantum Weyl algebra and its auxiliary
space ``$h$'', such that the trace over this auxiliary space of
the monodromy of $M$ operators gives $Q$-operator:
\begin{equation}\label{Mmonodromy}
\ds Q(x^2)\;=\;\tr_h\;\left(
M_{h,1}(x)\*M_{h,2}(x)\;\cdots\;M_{h,\nsites}(x)\right)\;.
\end{equation}
The commutativity of $Q$ with $T$ must provide
the intertwining relation for $M_{h,f}$ and $L_f$:
\begin{equation}\label{LLM}
\ds
\widetilde{L}_{h}(x/y)\,*\, L_{f}(x)\* M_{h,f}(y)\;=\;
M_{h,f}(y)\* L_{f}(x)\,*\, \widetilde{L}_{h}(x/y)\;,
\end{equation}
where ``$*$'' means the $2\times 2$ matrix multiplication,
and $\widetilde{L}_h(z)$ is an auxiliary $L$-operator.
Below we will give the explicit form of this
$\widetilde{L}$, but now we investigate the triangle
relations following from eq. (\ref{LLM}) and providing
Baxter's $TQ=Q'+Q''$ equation.

\subsection{Triangle relations}

Triangle relations should appear when we choose such $x/y$ in eq.
(\ref{LLM}) that $\widetilde{L}$ becomes degenerate as a $2\times 2$
matrix. Without lost of generality we may put $\ds\det\widetilde{L}(1)=0$.
We imply $\widetilde{L}^{-1}$ to be normalized to the determinant
of $\widetilde{L}$, i.e. $\widetilde{L}(1)$ and
$\widetilde{L}^{-1}(1)$ must be orthogonal. Thus introducing the
notations $\a$ and $\a^+$ for two appropriate elements for
$h$-algebra, one may write:
\begin{equation}\label{wtL-decomposition}
\ds
\widetilde{L}^{}_{h}(1)\;=\;
{-\a^+\choose 1}\;k_1\;(-\a,1)\;\;\;
\mbox{and}\;\;\;
\widetilde{L}^{-1}_{h}(1)\;=\;
{1\choose \a}\;k_2\;(1,\a^+)\;,
\end{equation}
Writing this decomposition, we do not impose no conditions on
$\a$, $\a^+$ and unknown factors $k_1$ and $k_2$.
The experience of usual Toda chain \cite {GP} says that when
$q\mapsto 1$, $\a$ and $\a^+$ become usual bosonic annihilation
and creation operators, this inspires our notations.
In the next derivations we will suggest the
invertibility of $\a$ and $\a^+$,
so that the decompositions (\ref{wtL-decomposition})
are written without loss of generality. Still we know
nothing about algebra of $\a$ and $\a^+$. To get something
applicable, let us introduce an element $\N$, such that
a sort of $q$-oscillator relations hold:
firstly, for any function $g$ let
\begin{equation}\label{a-N}
\ds \a\* g(\N)\;=\;g(\N+1)\*\a\;,
\end{equation}
and secondly let there exists a function $[\N]$:
\begin{equation}\label{aa+}
\ds
\a^+\*\a\;=\;[\N]\;,\;\;\;
\a\*\a^+\;=\;[\N+1]\;.
\end{equation}
Element $\N$ is introduced without loss of generality as a pair
to $\a$. Operators $\widetilde{L}$ must form an integrable chain,
this provides relations (\ref{aa+}), and therefore
$k_1\;=\;k_1(\N)$ and $k_2\;=\;k_2(\N)$ in eq.
(\ref{wtL-decomposition}). Thus, parameterization
(\ref{wtL-decomposition}) is the general one. The degenerate
$\widetilde{L}^\pm(1)$ matrices becomes the orthogonal projectors
when one choose the unknowns $k_1=1/(1+[\N+1])$ and
$k_2=1/(1+[\N])$.

Now let us write the explicit form of the triangle relations:
\begin{equation}\label{tr-1}
\ds\begin{array}{l}\ds (-\a,1)\,L(x)M(x)=M'(x)\,(-\a,1)\,,\\
\\
\ds L(x)M(x)\,{1\choose\a}={1\choose\a}\,M''(x)\;,
\end{array}
\end{equation}
and
\begin{equation}\label{tr-2}
\ds\begin{array}{l} \ds M(x)L(x)\,{-\a^+\choose 1}= {-\a^+\choose
1}\,\underline{M}'(x)\,,\\
\\
\ds (1,\a^+)\,M(x)L(x)\;=\;\underline{M}''(x)\,(1,\a^+)\,.
\end{array}
\end{equation}
Inserting the projectors into the product of $L(x)$ and $M(x)$ monodromies
and taking the traces, one obtains in usual way the consequence of eqs.
(\ref{tr-1}) of (\ref{tr-2}) Baxter's equation
\begin{equation}
\ds
T(x^2)\;Q(x^2)\;=\;\left((-\,x)^\nsites\,\prod_f\,\Uh_f\right)\;(Q'(x^2)\;
+\;Q''(x^2))\;,
\end{equation}
where extra multiplier appears due to our normalization of the transfer
matrix,(see definition (\ref{T-matrix})).
Note, both (\ref{tr-1}) and (\ref{tr-2}) have to provide the same Baxter's
equation, hence the traces of $M^\#$ and $\underline{M}^\#$ monodromies
should  be the same.

The spectral parameter $x$ in $L$ operator (\ref{L-operator})always stay
in the combination $x\,\Uh$ therefore the shift of the spectral parameter
thus may appear as
\begin{equation}
\ds
g(q^{1/2}x\,\Uh_f)\;\equiv\;\Vh_f^{-1}\;g(x\,\Uh_f)\;
\Vh_f^{}\;.
\end{equation}
Due to this property we can put $x=1$ for the shortness and omit the
spectral parameter in our formulae, the $x$ may be restored subsequently
in all equations by shift $\Uh_f\;\mapsto\;x\,\Uh_f$.

The triangle equations (\ref{tr-1},\ref{tr-2}) are equivalent
to two systems:
\begin{equation}\label{tr-11}
\ds\left\{\begin{array}{clc}
&\ds M'\;=\;-\;\a\,\Vh\;M\;,\;\;\;
M''\;=\;q^{-1/2}\,\lambda\;(\a\,\Vh)^{-1}\;M\;,&\\
&&\\
&\ds M\;\a\;=\;\Vh^{-1}\;
(\Uh^{-1}\;-\;\Uh\;+\;q^{-1/2}\,\lambda\;\a^{-1}\;\Vh^{-1})\;M\;,
\end{array}\right.\end{equation}
and
\begin{equation}\label{tr-22}
\ds\left\{\begin{array}{clc}
&\ds \underline{M}'\;=\;-\,M\;q^{-1/2}\,\lambda\,\Vh^{-1}\;\a^{+}\;,\;\;\;
\underline{M}''\;=\;M\;\Vh\,(\a^{+})^{-1}\;,&\\
&&\\
&\ds-\,q^{-1/2}\lambda\;\a^+\;M\;=\;M\;
(\Uh\;-\;\Uh^{-1}\;-\;(\a^{+})^{-1}\,\Vh)\;\Vh\;.&
\end{array}\right.\end{equation}
In Baxter's equation it is implied
$Q'(x^2)\sim Q(q^{-1}x^2)$ and $Q''(x^2)\sim Q(q\,x^2)$
up to some operator-valued multipliers. These multipliers
are to be integrals of motion in a form of pure product over $f$.
There is only one such integral of motion, it is $t_0$, and hence
there must exist a monomial function $\phi(\Uh)\sim\Uh^c$ such that
\begin{equation}\label{Mprime}
\ds
M'\;=\;\Vh\;M\;\phi(\Uh)\;\Vh^{-1}\;,\;\;\;
\underline{M}'\;=\;\Vh\;\phi(\Uh)\;M\;\Vh^{-1}\;,
\end{equation}
and therefore
\begin{equation}
\ds
M''\;=\;-q^{-1/2}\,\lambda\,\Vh^{-1}\;M\;\Vh\,\phi^{-1}(\Uh)\;,
\;\;\;
\underline{M}''\;=\;-q^{-1/2}\,\lambda\,
\phi^{-1}(\Uh)\,\Vh^{-1}\;M\;\Vh\;.
\end{equation}
The same $\phi(\Uh)$ is used for $M$ and $\underline{M}$ because
of $Q$ must be the same. It is important that in equations
(\ref{Mprime}) the multiplier $\phi(\Uh)$ stands from the right of to $M$
for $M'$ and from the left of $M$ for $\underline{M}'$. The order of
multipliers is governed by Yang-Baxter equation (\ref{LLM}).

In general one may put $\phi(\Uh)$ to the other sides, this would
give another system for $M$ with another solution. We will not
investigate such case separately, because of there exists an
involutive automorphism $\tau$, defined as
\begin{equation}\label{tau-involution}
\ds \Vh^\tau\;=\;\Vh\;,\;\;\; \Uh^\tau\;=\;\Uh^{-1}\;,\;\;\;
q^\tau\;=\;q\;,
\end{equation}
such that $L$-operator is invariant with respect to
$\tau$-involution:
\begin{equation}
\ds L(1)^\tau\;=\;-\sigma_3\;L(1)\;\sigma_3\;.
\end{equation}
Also important is that $\tau$ does not change $q$. Therefore $\ds
T^\tau(x^2)=(-)^\nsites\,T(x^2)$, and the another case of
positions of $\phi$ just corresponds to the consideration of
$M^\tau$.

With these expressions for $M'$ and $M''$
systems (\ref{tr-11},\ref{tr-22})
are equivalent to
\begin{equation}\label{tr-all}
\ds\begin{array}{lcl}
(i)   &:& \ds -\,\a\;M\;=\;M\;\phi(\Uh)\,\Vh^{-1}\;,\\&&\\
(ii)  &:& \ds -q^{-1/2}\lambda\,M\;\a^+\;=\;
\Vh\;\phi(\Uh)\;M\;,\\&&\\
(iii) &:& \ds M\;\a\;=\;
\Vh^{-1}\,\left(\Uh^{-1}\,-\Uh\,+\,
q^{-1/2}\lambda\,\a^{-1}\,\Vh^{-1}\right)\;
M\;,\\&&\\
(iv)  &:& \ds -q^{-1/2}\lambda\,\a^+\;M\;=\;M\;
\left(\Uh\,-\,\Uh^{-1}\,-\,(\a^+)^{-1}\,\Vh\right)\,\Vh\;.
\end{array}\end{equation}
It is useful to complement system (\ref{tr-all})
by equations with $k_1,k_2$ following from (\ref{LLM}):
\begin{equation}
\ds\begin{array}{lcl}
(v)   &:& \ds M\;\phi^{-1}(\Uh)\,k_1(\N)\;=\;
\phi^{-1}(\Uh)\,k_1(\N)\;M\;,\\&&\\
(vi)  &:& \ds M\;\phi^{-1}(q^{-1/2}\Uh)\,k_2(\N)\;=\;
\phi^{-1}(q^{-1/2}\Uh)\,k_2(\N)\;M\;.
\end{array}
\end{equation}
This is the final set of equations that we are going to solve. We
will give the solution of it in two forms. The fist one is a
formal series solution that admits an interpretation of $\a$ and
$\a^+$ as $q$-oscillator (spectrum of $\N$ is non-negative
integers, and there exists the vacuum vector for $\a$). Another
form implies the Weyl algebra parameterization of $\a$, $\a^+$,
when the permutation between $h$ and $f$ spaces pays the
significant role. Actually these two forms differ by the notion
of the trace in $h$-space, and $q$-oscillator trace will give
$Q_-$ while the Weyl trace will give $Q_+$.

\subsection{Series solution}

First, let us test system (\ref{tr-all}) for the formal
operator arguments of $M$. Just considering the expressions
$M\,\x\,M^{-1}\,\x^{-1}$ for several $\x$, one may conclude
\begin{equation}
\ds M\;=\;M(\a\,\Vh\,,\,\Uh\,,\,\N)\;.
\end{equation}
Hence
\begin{equation}\label{center}
\ds
M\* q^{\N}\,\Uh^2\;=\;\Uh^2\,q^{\N}\*M\;,
\end{equation}
this trivializes two equivalent relations
$(v)$ and $(vi)$ of system (\ref{tr-all}).

The further analysis of (\ref{tr-all}) we start from the
permutation-like relation (i). The relations like
\begin{equation}
\ds \x\* M\;=\;M\*\y
\end{equation}
are to be solved as
\begin{equation}
\ds
M\;=\;\sum_{n\in\mathbb{Z}}\;\;\x^n\* G \* \y^{-n}\;,
\end{equation}
and in the case of (\ref{tr-all}-$(i)$) this gives
\begin{equation}\label{ansatz-M}
\ds M\;=\;\sum_{n\in\mathbb{Z}}\;\; \a^n\* G(\N,\Uh^2)\*
\left(-\,\Vh\,\phi^{-1}(\Uh)\right)^n\;.
\end{equation}
$G$ does not depend on $\a\Vh$, because any such dependence
may be extracted to $\a^n$. Now all other relations from
(\ref{tr-all}) must give recursion relations for $G$.
Equation $(iii)$ of (\ref{tr-all}) is equivalent to
\begin{equation}\label{bessel-relation}
\ds\begin{array}{l} \ds (u-u^{-1})\,G(N,u^2)\,=\,\\
\\
\ds=\; -q^{-1/2}\,\lambda\,\phi^{-1}(u)\,G(N,qu^2)\,+\,
\phi(q^{-1/2}u)\,G(N-1,q^{-1}u^2)\,.
\end{array}
\end{equation}
Eq. $(ii)$ of (\ref{tr-all}) gives another permutation-like
structure, but with the formal correspondence
$\ds \a^+\;=\;[\N]\;\a^{-1}$ it gives
\begin{equation}\label{factorial-relation}
\ds
{G(N-1,u^2)\over G(N,q u^2)}\;=\;q^{1/2}\lambda\;
{[N]\over\phi^2(u)}\;.
\end{equation}
Due to (\ref{factorial-relation})
$M$ may be rewritten in the form of the other permutation-like structure:
\begin{equation}\label{ansatz-M+}
\ds M\;=\;\sum_{n\in\mathbb{Z}}\;\;
\left(-q^{1/2}\lambda^{-1}\,\Vh\,\phi(\Uh)\right)^n\*
G(\N,\Uh^2)\* (\a^{+})^{-n}\;.
\end{equation}
Moreover, this allows one
to write $M$  without negative powers of $\a$ or $\a^+$:
\begin{equation}\label{q-osc-M}
\ds\begin{array}{ccl}
\ds M & = & \ds G(\N,\Uh^2)\;+\\
&&\\
&+&\ds \sum_{n=1}^\infty\;
\a^n\* G(\N,\Uh^2)\* \left(-\,\Vh\,\phi^{-1}(\Uh)\right)^n\;+\\
&&\\
&+&\ds \sum_{n=1}^\infty\;
\left(-q^{1/2}\lambda^{-1}\,
\Vh\,\phi(\Uh)\right)^{-n}\* G(\N,\Uh^2)\*
(\a^{+})^{n}\;.
\end{array}
\end{equation}
Apparently, this form is good for $q$-oscillator representation.

The equation $(iv)$ of (\ref{tr-all}) coincides with
(\ref{bessel-relation})
if one uses the series (\ref{ansatz-M+}). But it is important to
note that in general (\ref{bessel-relation}) and
(\ref{factorial-relation}) are not compatible. Their
compatibility (i.e. zero curvature) condition is the following
functional relation for $\phi(\Uh)$ and $[\N]$:
\begin{equation}\label{compat}
\ds\begin{array}{clc}
&\ds q^{-1/2}\lambda
\left({[\N]\over\phi(q^{1/2}\Uh)}-{[\N-1]\over\phi(q^{-1/2}\Uh)}\right)
\;=\;&\\&&\\
&\ds =\;\Uh^{-1}
\left(1-q^{-1/2}{\phi(\Uh)\over\phi(q^{1/2}\Uh)}\right)
\;-\;\Uh
\left(1-q^{1/2}{\phi(\Uh)\over\phi(q^{1/2}\Uh)}\right)\;.
\end{array}
\end{equation}
Here we used that $\phi(\Uh)\sim\Uh^c$.

Eq. (\ref{compat}) has only two solutions for
$\phi(\Uh)$ and $[\N]$,  corresponding
to $|q|\,<\,1$ and $|q|\,>\,1$. In our case $|q|\,<\,1$
\begin{equation}
\ds\phi(u)\;=\;-q^{-1/2}\,\alpha\,u^{-1}\;,\;\;\;
[N]\;=\;-q^{1/2}\,{\alpha\over\lambda}\;(1\;-\;q^{-N})\;,
\end{equation}
where $\alpha$ is a complex parameter, $[\N]$ is normalized so as $[0]=0$.
With these $\phi(u)$ and $[N]$ eqs. (\ref{bessel-relation}) and
(\ref{factorial-relation}) may be solved easily giving
\begin{equation}
\ds
G_{|q|<1}(N,u^2)\;=\;\g_{\alpha,\lambda/\alpha}(N,N-s)\;,
\end{equation}
where $u^2\,\equiv\,q^s$ (see eq. (\ref{xsf})), and
$\g_{\alpha,\beta}(n,m)$ is defined by (\ref{g-function}).
Parameter $\alpha$ is an avoidable scale of
$\Uh$ and it is convenient
to put it to unity, $\alpha\equiv 1$.
Note that expressions like $(x;q)_\infty$ in $\g$-function
appear as the appropriate solutions of difference relations
\begin{equation}
\ds (x;q)_\infty\;=\;(1-x)\;(qx;q)_\infty\;,
\end{equation}
and separation between $|q|<1$ and $|q|>1$ is originated
from the unavoidable sign
of the quadratic exponent $q^{\pm N(N-s)}$
\footnote{
The other solution of zero curvature condition is
\begin{equation}
\ds\phi(u)\;=\;\alpha\;u\;,\;\;\;
[N]\;=\;-{\alpha\over\lambda}\;
(1\;-\;q^N)\;.\nonumber
\end{equation}
This gives
\begin{equation}
\ds\begin{array}{l} \ds  G(N,q^s)=q^{-\,N(N-s)}
\left(q^{-1/2}{\lambda\over\alpha}\right)^{N-s}
\left(q^{-1/2}\,\alpha\right)^{N}\times\\
\\
\ds\times  {\left(q^{-1-N+s};q^{-1}\right)_\infty\,
\left(q^{-1-N};q^{-1}\right)_\infty\over (q^{-1};q^{-1})_\infty}.\nonumber
\end{array}
\end{equation}
}.
With $\phi(\Uh)$ defined, the final expressions for $M$ are:
the short two
\begin{equation}\label{Mfinal}
\ds\begin{array}{l} \ds M\;=\;\sum_{n\in\mathbb{Z}}\;
\a^n\;\g_{1,\lambda}(\N,\N-\s)\;(\Uh\Vh)^n\;\equiv\;\\
\\
\ds\equiv \sum_{n\in\mathbb{Z}}\; (\lambda\Uh\Vh^{-1})^{-n}\;
\g_{1,\lambda}(\N,\N-\s)\;(\a^+)^{-n}\;,
\end{array}
\end{equation}
and $q$-oscillator-type
\begin{equation}\label{Mfinal-qosc}
\ds\begin{array}{ccl}
\ds M & = & \g_{1,\lambda}(\N,\N-\s)\;+\\
&&\\
\ds &+& \ds \sum_{n=1}^\infty\;
\a^n\;\g_{1,\lambda}(\N,\N-\s)\;(\Uh\Vh)^n\;+\\
&&\\
&+&\ds\sum_{n=1}^\infty\;
(\lambda\Uh\Vh^{-1})^{n}\;
\g_{1,\lambda}(\N,\N-\s)\;(\a^+)^{n}\;,
\end{array}
\end{equation}
where, recall eq. (\ref{xsf}), $\Uh^2\;=\;q^{\s}$.

Substituting $\phi(\Uh)=-q^{-1/2}\Uh^{-1}$ into
the expressions for $M'$ and $M''$, (\ref{Mprime}),
and using our definition of the transfer matrix (\ref{T-matrix}),
we obtain the Baxter equation exactly in the form
(\ref{Baxter}).

Existence of the form (\ref{Mfinal-qosc}) allows one to interpret
$\a,\a^+$ exactly as $q$-oscillator generators, such that the
spectrum of $\N$ is $0,1,2,...$ (we have normalized $[\N]$ so
that $[0]\;=\;0$), and the state $|\N=0>$ is the vacuum,
$\a\,|\N=0>=0$. Thus one may define the $q$-oscillator trace of
any operator $F\;=\;F(\a,\a^+,\N)$,
\begin{equation}
\ds
F\;=\;f_0^{}(\N)\;+\;\sum_{n\geq 1}\;f_n^{}(\N)\,\a^n\;+\;
\sum_{n\geq 1}\;f_n^+(\N)\,(\a^+)^n\;,
\end{equation}
taking such trace one has to take $\a^0$ and $(\a^+)^0$-th
components and then take the sum over $\N=0,1,2,...$:
\begin{equation}
\ds \tr_{q-osc}\;F(\a,\a^+,\N)\;\stackrel{def}{=}\; \sum_{n\geq
0}\;f_0^{}(n)\;.
\end{equation}
Being applied to the monodromy (\ref{Mmonodromy})
of $M$ (\ref{Mfinal}), this trace definition gives
exactly $Q_-(x^2)$, eq. (\ref{Q-}).

In general one may obtain $Q_+$ immediately, considering the
$\tau$ - involution applied to $M$ and to $Q_-$:
\begin{equation}
\ds
M^\tau\;=\;\sum_{n\in\mathbb{Z}}\;\left(\a\,\Vh\Uh^{-1}\right)^n\;
\g_{1,\lambda}(\N,\N+\s)\;.
\end{equation}
But there are two objections to consider this case: firstly,
$\tau$-involution changes a little the Baxter equation, and
secondly, $M^\tau$ is the degenerate operator,
\begin{equation}
\ds (\a\Vh\Uh^{-1}-1)\* M^\tau\;=\;0\;,
\end{equation}
and hence we will look for another way to obtain $Q_+$ operator.

\subsection{Extraction of a permutation}

Solving eqs. (\ref{tr-all}), we mentioned the permutation-like
relations. In this section let us suppose that the quantum space
$f$ and the auxiliary one $h$ are isomorphic. Our aim is to
extract the permutation operator, giving eq. (i) of
(\ref{tr-all}) ``by hands''. As previously, we deal with the case
$|q|\;<\;1$, $x\;=\;1$, $\Uh^2\;=\;q^{\s}$, and
\begin{equation}\label{phi-n-recall}
\ds
\phi(\Uh)\;=\;-\,q^{-1/2}\,\Uh^{-1},\;\;\;
[\N]\;=\;-\,q^{1/2}\,\lambda^{-1}\,(1\;-\;q^{-\N})\;,
\end{equation}
so that we are looking for another realization of the same
operator $M$. We will search for $M$ in the form
\begin{equation}\label{MmP}
\ds M\;=\;\mathcal{M}\* P_{h,f}\;,
\end{equation}
where
\begin{equation}\label{permutation}
\ds
\a\;P_{h,f}\;=\;P_{h,f}\;(\Uh\Vh)^{-1}\;,\;\;\;
\N\;P_{h,f}\;=\;P_{h,f}\;\s\;,\;\;\;
P_{h,f}^2\;=\;1\;.
\end{equation}
Here the first relation is exactly eq. (i) of (\ref{tr-all}), the
second one is the consequence of eq. (\ref{center}), and the last
one is the definition of the permutation. System (\ref{tr-all})
for operator $\mathcal{M}$ can be rewritten as follows:
\begin{equation}\label{tr-all-m}
\ds\begin{array}{ccl}
\ds (i) & : & \ds
\a\*\mathcal{M}\;=\;\mathcal{M}\*\a\;,\\
&&\\
\ds (ii) & : & \ds
\Uh^{-1}\Vh\*\mathcal{M}\;=\;\mathcal{M}\*(1\,-\,\Uh^2)\,\Uh^{-1}\Vh\;,\\
&&\\
\ds (iii) & : & \ds \mathcal{M}\*(\Uh\Vh)^{-1}\;=\;
(\Uh\Vh)^{-1}\;\left(1\,-\,\Uh^{2}\,+\,q^{-1}\lambda\,
\a^{-1}\Vh^{-1}\Uh\right)\*\mathcal{M}\;,\\
&&\\
\ds (iv) & : & \ds q^{-\N}\*\mathcal{M}\;=\;
\mathcal{M}\*\left(1\,+\,q^{-1}\lambda\,
(1-q\Uh^2)^{-1}\,\a^{-1}\Vh^{-1}\Uh\right)\;q^{-\N}\;.
\end{array}\end{equation}
Solution of it is given by
\begin{equation}\label{m}
\ds \mathcal{M}\;=\;
(-\lambda\a^{-1}\Vh^{-1}\Uh;q)_\infty\;(q\,\Uh^2;q)_\infty\;.
\end{equation}
Operator (\ref{MmP}) with the definitions
(\ref{permutation},\ref{m}) does solve the  system of the relations
(\ref{tr-all}). Using the series decomposition for the compact
quantum dilogarithms, one may obtain the series representation
\footnote{ There is used
\begin{equation}
\ds
(\Vh^{-1}\Uh)^n\;=\;q^{n(n+1)/2}\;\Uh^{2\,n}\;(\Uh\Vh)^{-n}\;.\nonumber
\end{equation}
}
for $M$ (\ref{MmP}):
\begin{equation}
\ds
M\;=\;\sum_{n\geq 0}\;
{q^{n^2}\over (q;q)_n}\;
\lambda^n\,\Uh^{2\,n}\;(q^{1+n}\Uh^2;q)_\infty\;
(\Uh\Vh)^{-n}\;P_{h,f}\;(\Uh\Vh)^n\;.
\end{equation}
Note, function $\g$, eq. (\ref{g-function}), appears in this
decomposition:
\begin{equation}\label{M-g-per}
\ds
M\;=\;\sum_{n\geq 0}\;
\g_{1,\lambda}(n+\s,n)\;(\Uh\Vh)^{-n}\;P_{h,f}\;(\Uh\Vh)^n\;.
\end{equation}
In this form all the $h$-space operators $\a,\a^+$ and $\N$ are hidden
into the permutation symbol.
The permutation operator allows one to calculate the trace in the
auxiliary space $h$ in the invariant way via
\begin{equation}
\ds
\tr_{inv}\;\left(P_{h,1}\,P_{h,2}\,\cdots\,P_{h,\nsites}\right)
\;=\;P\;,
\end{equation}
where $P$ is the cyclic shift operator for the chain
$f=1,2,...,\nsites,\nsites+1\sim 1$:
\begin{equation}
\ds \Uh_f\;P\;=\;P\;\Uh_{f+1}\;,\;\;\;
\Vh_f\;P\;=\;P\;\Vh_{f+1}\;,\;\;\;\;f\sim f+\nsites\;.
\end{equation}
The shift is one of the integrals of motion.
Now using eq. (\ref{M-g-per}) and the definition of the shift operator,
one obtains exactly $Q_+$ (\ref{Q+}) for the trace of $M$-monodromy up to
the shift:
\begin{equation}
\ds
Q_+\;P\;=\;P\;Q_+\;=\;\tr_{inv}\;\left(M_1\;M_2\;\cdots\;M_{\nsites}\right
)\;.
\end{equation}

Now both forms of $M$-operators have been obtained, eqs. (\ref{q-osc-M})
and (\ref{M-g-per}), actually coincide. To show it, let us represent
$P_{h,f}$ in the following form:
\begin{equation}
\ds P_{h,f}\;=\;\sum_{n\in\mathbb{Z}}\;
\delta(\N-\s\;=\;n)\;(\a\Uh\Vh)^n\;,
\end{equation}
where $\delta(\N-\s=n)$ is the projector of $\N-\s$ into a state with the
eigenvalue
$n$:
\begin{equation}
\ds
(\N-\s)\;\delta(\N-\s=n)\;=\;\delta(\N-\s=n)\;(\N-\s)\;=\;
n\,\delta(\N-\s=n)\;.
\end{equation}
With this form of $P_{h,f}$ eq. (\ref{M-g-per}) could be written as
follows:
\begin{equation}
\ds M\;=\;\sum_{n,k}\;
\a^k\,\g_{1,\lambda}(n+\s,n)\,\delta(\N-\s=n)\;(\Uh\Vh)^k\;.
\end{equation}
Now one may take the sum over $n$ using the projectors as the delta
symbols,
and exactly eq. (\ref{Mfinal}) appears:
\begin{equation}
\ds
M\;=\;\sum_{k}\;\a^k\,\g_{1,\lambda}(\N,\N-\s)\;(\Uh\Vh)^k\;.
\end{equation}

Such exercises with the projector decomposition of operators are
rather formal.One may consider projectors and spectral
decompositions of many types, imposing some extra conditions for
the spectra of the operators involved. What is actually the
difference between both  $Q$ operators: the difference is
the conjecture about the spectrum of $\N$. Due to the Weyl
algebra relations, the spectrum of $\N$ must be equidistant,
\begin{equation}
\ds\N\;\in\; Z\;+\;\zeta\;.
\end{equation}
In the case when $\zeta=0$ we get q-oscillator representation.
In the case when $\zeta$ is the same as for
$\s\in\mathbb{Z}+\zeta$, we get the isomorphism between $h$
and $f$ spaces and the permutation extracted representation. In
general one may generalize both $Q_+$ and $Q_-$ into $Q_\zeta$,
dealing with arbitrary characteristics of $\N$.
\begin{equation}\label{Qzeta}
\ds Q_\zeta\;=\;\sum_{\{n_f\in\mathbb{Z}+\zeta\}}\;
\left(\prod_f\;\g_{1,\lambda}(n_f,n_f-\s_f)\right)\*
\left(\prod_f\;(\Uh\Vh)_f^{n_f-n_{f-1}}\right)\;.
\end{equation}
A summation over $n\in\mathbb{Z}$ may be restricted in
q-hypergeometry by the factor
\begin{equation}
\ds{1\over (q;q)_n}\;=\;{(q^{1+n};q)_\infty\over
(q;q)_\infty}\;=\;0\;\;\;\mbox{for}\;\;\;n<0\;.
\end{equation}
Such restrictions in eq. (\ref{Qzeta}) appear when $\zeta=0$ and when
$\s\in
Z+\zeta$, these are exactly the cases of $Q_-$ and $Q_+\,P$.

Similarly to the spectral decomposition of the permutation
operator, one may write down the spectral decomposition of the
shift operator:
\begin{equation}
\ds P\;=\;\sum_{\{n_f\in\mathbb{Z}\}}\;
\left(\prod_f\;\delta(\s_f=n_f+\zeta)\right)\;
\left(\prod_f\;(\Uh\Vh)_f^{n_f-n_{f-1}}\right)\;.
\end{equation}
In this formula it is implied that $\zeta$  is the
characteristics of $\s_f$.  An example of application of such
formula, i. e. explicit extraction of the shift operator, is the
following summation, where the shift $n_f\mapsto n_f+\s_f$ is
done:
\begin{equation}\label{P-trick}
\ds\begin{array}{cl} &\ds \sum_{\{n_f\in\zeta+\mathbb{Z}\}}\,
G(\{n_f,n_f-\s_f\})\;\prod_{f}\,(\Uh\Vh)_f^{n_f-n_{f-1}}\;=\\
&\\
&\ds=\; \sum_{\{n_f\in\mathbb{Z}\}}\; G(\{n_f+\s_f,n_f\})\;
\prod_{f}\;(\Uh\Vh)_f^{n_{f+1}-n_f}\* P\;.
\end{array}\end{equation}
To obtain it, one has to apply the spectral decomposition of each
$\s_f$, and then make the re-summation. This trick gives
$Q_\zeta\;=\;Q_+(x^2)\;P$ when $\s_f\in\zeta+\mathbb{Z}$.

\section{Properties of $M$ operators}

\subsection{Auxiliary $L$-operator}

\begin{prop}
Equation (\ref{LLM}), provided by eqs.
(\ref{tr-all},\ref{center},\ref{phi-n-recall}), holds for
\begin{equation}\label{wtL}
\ds
\widetilde{L}(x)\;=\;\left(\begin{array}{ccc}
\ds x\,q^{\N/2}\,-\,x^{-1}q^{-\N/2} &,& \ds\lambda\,\a^+q^{\N/2} \\
&&\\
\ds \lambda\, q^{\N/2}\a &,& -\,\lambda\, x^{-1}
q^{\N/2}\end{array}\right)\;.
\end{equation}
\end{prop}
To be exact, in our normalization $M=M(1)$, for which eqs.
(\ref{tr-all}, \ref{center}, \ref{phi-n-recall}) are written down,
eq. (\ref{RLL}) looks like
\begin{equation}
\ds M\* L(x) \* \widetilde{L}(x)\;=\;\widetilde{L}(x)\* L(x)\* M\;,
\end{equation}
and $M$ must intertwine each power of $x$. \footnote{ Useful
relations following from eqs. (\ref{tr-all}) are
\begin{equation}
\ds M\*(\Uh\Vh)^{-1}\;=\;\a\*M\;,\;\;\;
M\*\Uh\Vh^{-1}\;=\;q^{\N}\,(\lambda\Uh\Vh^{-1}+q^{1/2}\a)\*M\;,\nonumber
\end{equation}
and
\begin{equation}
\ds
\Vh\Uh^{-1}\* M\;=\;M\*\lambda\,\a^+\;,\;\;\;
\Uh\Vh \*M\;=\;M\*\lambda\,q^{\N}(\Uh\Vh+q^{-1/2}\a^+)\;.\nonumber
\end{equation}
} Note, as far the quantum Lax operator (\ref{L-operator}) is
called ``the relativistic Toda chain $L$-operator'', then operator
(\ref{wtL}) is to be called ``the relativistic Dual self-trapping
$L$-operator'', see e.g. \cite{Sklyanin3}.

\subsection{Intertwining}

Now let us consider the commutation relations of different $Q$-operators.
Let the operators $Q_1(y)$ and $Q_2(x)$ are constructed with the help of
different local $M_{h_1,f}(y)$ and $M_{h_2,f}(x)$ (here we imply
different characteristics of $h_1$ and $h_2$).

\begin{prop}
Two products: $M_{h_1,f}(y)\*M_{h_2,f}(x)$ and
$M_{h_2,f}(x)\*M_{h_1,f}(y)$, are connected by a canonical
mapping $\ds K_{h_1,h_2}(y/x)$ of the pair of Weyl algebras $h_1$
and $h_2$:
\begin{equation}
\ds K_{h_1,h_2}\left({y\over x}\right)\;
M_{h_1,f}(y)\;M_{h_2,f}(x)\;=\; M_{h_2,f}(x)\;M_{h_1,f}(y)\;
K_{h_1,h_2}\left({y\over x}\right)\;,
\end{equation}
where $K$ acts as follows:
\begin{equation}
\ds\begin{array}{ll} \ds K(z)\a_{1}^{+}  =  \ds
z^{-1}\a_2^+K(z),& \ds K(z)q^{\N_1}  =  \ds
{1+q^{1/2}z\a_1^{}\a_2^+\over
 1+q^{1/2}z^{-1}\a_1^{}\a_2^+}q^{\N_2}K(z),\\
&\\
\ds K(z)\a_{2}^{} = \ds z\a_{1}^{}K(z),& \ds K(z)q^{\N_2} = \ds
q^{\N_1} {1+q^{1/2}z^{-1}\a_1^{}\a_2^+\over
 1+q^{1/2}z\a_1^{}\a_2^+}K(z).
\end{array}\end{equation}
\end{prop}
As an example we give the realization of $K(z)$ with the permutation
extracted:
\begin{equation}
\ds
K_{h_1,h_2}(z)\;=\;
{\left(-\,q^{1/2}\;\,z\;\;\a_1^{}\,\a_2^{+}\,;\,q\right)\over
\left(-\,q^{1/2}\,z^{-1}\a_1^{}\,\a_2^{+}\,;\,q\right)}\;
z^{-\N_1-\N_2}\;P_{h_1,h_2}\;,
\end{equation}
where $P_{h_1,h_2}$ - usual external permutation of the spaces $h_1$ and
$h_2$.
This permutation may be canceled from $KMM$ equation, and
following relation for the $Q$-monodromies appears:
\begin{equation}\label{pseudocom}
\ds \check{K}_{h_1,h_2}\left({y\over x}\right)\;
\widehat{Q}_{h_1}(y^2)\;\widehat{Q}_{h_2}(x^2)\;=\;
\widehat{Q}_{h_1}(x^2)\;\widehat{Q}_{h_2}(y^2)\;
\check{K}_{h_1,h_2}\left({y\over x}\right)\;,
\end{equation}
where
\begin{equation}
\ds \check{K}_{h_1 h_2}\left({y\over x}\right)= P_{h_1 h_2}K_{h_1
h_2}\left({y\over x}\right)= \left({x\over y}\right)^{\N_1+\N_2}
{\left(-q^{1/2}{y\over x}\a_1^+\a_2^{};q\right)_\infty\over
\left(-q^{1/2}{x\over y}\a_1^+\a_2^{};q\right)_\infty},
\end{equation}
and  $\widehat{Q}$ -- the monodromy of $M$ operators, $\ds
Q(x^2)\;=\;\tr_h\;\widehat{Q}_h$. Eq. (\ref{pseudocom}) leads to
the pseudo-commutation of the pair of $Q$ matrices with {\em
different} $\zeta$-characteristics and allows one to calculate
the wronskian.

\subsection{Wronskian}

To calculate the wronskian, it is necessary to consider eq.
(\ref{pseudocom}) with $\ds {x\over y}\;=\;q^{1/2}$. Then
$\ds\check{K}(q^{-1/2})\;=\;q^{(\N_1+\N_2)/2}\,(1+\a_1^+\a_2^{})$,
and
\begin{equation}
\ds\begin{array}{clc} &\ds q^{(\N_1+\N_2)/2}\,(1+\a_1^+\a_2^{})\;
\widehat{Q}_{1}(q^{-1}x^2)\;\widehat{Q}_{2}(x^2)\;=&\\
&&\\
&\ds =\; \widehat{Q}_{1}(x^2)\;\widehat{Q}_{2}(q^{-1}x^2)\;
(1+\a_1^+\a_2^{})\,q^{(\N_1+\N_2)/2}\;.&
\end{array}\end{equation}
Let $\delta_{W}$ be a projector to the subspace
$\a_1^+\,\a_2^{}\;=\;-1$, i.e.
\begin{equation}
\ds\delta_{W}\*(\a_1^+\a_2^{}\;+\;1)\;=\;(\a_1^+\a_2+1)\*\delta_W\;=\;0\;.
\end{equation}
Then the pseudo-commutation relation provides the following
triangle structure:
\begin{equation}
\ds\begin{array}{lcl}\label{dQQ}
\ds\widehat{Q}_1(q^{-1}x^2)\;\widehat{Q}_2(x^2)\;\delta_W & = &
\ds
\delta_{W}\;\widehat{Q}_1(q^{-1}x^2)\;\widehat{Q}_2(x^2)\;\delta_W\;,\\
&&\\
\ds\delta_{W}\;\widehat{Q}_1(x^2)\;\widehat{Q}_2(q^{-1}x^2) & = &
\ds
\delta_{W}\;\widehat{Q}_1(x^2)\;\widehat{Q}_2(q^{-1}x^2)\;\delta_W\;.
\end{array}\end{equation}
Locally we consider products:
\begin{equation}
\ds\begin{array}{ccl}
\ds M_1(y)\;M_2(x)\;\delta_W & = & \ds
\sum_{n\in\mathbb{Z}}\;\left(\lambda y\Uh\Vh^{-1}\right)^n\;
\mathcal{F}_{y,x}(\N_1,\N_2,\Uh^2)\;(\a_1^+)^n\;\delta_W\;,\\
&&\\
\ds\delta_W\;M_1(x)\;M_2(y) & = & \ds\delta_W\;
\sum_{m\in\mathbb{Z}}\;\a_2^m\;
\widetilde{\mathcal{F}}_{x,y}(N_1,N_2,\Uh^2)\;\left(y\Uh\Vh\right)^n\;,
\end{array}
\end{equation}
where the sum simplified due to
\begin{equation}
\ds \delta_{W}\*(\a_1^+)^n\,\a_2^{m-n}\;\equiv\;\delta_{W}\*
(-)^n\;\a_2^{m}\;.
\end{equation}
Triangle structure means that when $y^2=q^{-1}x^2$, both
$\mathcal{F}$ and $\widetilde{\mathcal{F}}$ depend actually only
on $\N_1+\N_2$ and $\Uh^2$.

For $x$ and $y$ arbitrary, one has
\begin{equation}
\ds \begin{array}{l} \ds \mathcal{F}_{y,x}(\N_1,\N_2,\Uh^2)
\stackrel{def}{=}\\
\\
\ds= \sum_{m\in\mathbb{Z}}(-\lambda x y \Uh^2)^m q^{m^2/2}
\g_{1,\lambda}(\N_1,\N_1-\s_y-m)\,\g_{1,\lambda}(\N_2+m,\N_2-\s_x),\\
\\
\ds \widetilde{\mathcal{F}}_{x,y}(\N_1,\N_2,\Uh^2)
\stackrel{def}{=}\\
\\
\ds= \sum_{n\in\mathbb{Z}}\;(-\lambda x y \Uh^2)^n\;q^{n^2/2}\;
\g_{1,\lambda}(\N_1+n,\N_1-\s_x)\,\g_{1,\lambda}(\N_2,\N_2-\s_y-n).
\end{array}\end{equation}
Here
\begin{equation}
\ds q^{\s_x}\;=\;x^2\,\Uh^2\;,\;\;\; q^{\s_y}\;=\;y^2\,\Uh^2\;.
\end{equation}
One may see,
\begin{equation}
\ds\widetilde{\mathcal{F}}_{x,y}(\N_1,\N_2,\Uh^2)\;\equiv\;
\mathcal{F}_{y,x}(\N_2,\N_1,\Uh^2)\;.
\end{equation}
 These sums may be calculated with the help of the
Rogers-Ramanujan summation formula. Auxiliary relations for this
calculations are:
\begin{equation}
\ds (\lambda x\Uh\Vh^{-1})^n\;(y\Uh\Vh)^n\;=\;
q^{n^2/2}\;(\lambda x y \Uh^2)^n\;,
\end{equation}
\begin{equation}
\ds{\g_{1,\lambda}(N+n,N-s)\over\g_{1,\lambda}(N,N-s)}\;=\;
q^{n\,(N-s)}\;{1\over (q^{1+N};q)_n}\;,
\end{equation}
\begin{equation}
{\g_{1,\lambda}(N,N-s-n)\over \g_{1,\lambda}(N,N-s)}\;=\;
q^{-n^2/2+n/2}\,(-\,\lambda\,q^s)^{-n}\,
(q^{s-N};q)_n\;,
\end{equation}
and the Rogers-Ramanujan celebrated identity is
\begin{equation}
\ds\begin{array}{l}
\ds\phantom{i}_1\Psi_1(x,y;z)\;\stackrel{def}{=}\; \ds
\sum_{n\in\,\mathbb{Z}}\;{(x;q)_n\over (y;q)_n}\,z^n\;=\;\\
\\
\ds=\; {(q;q)_\infty (y/x;q)_\infty (xz;q)_\infty
(q/xz;q)_\infty\over (y;q)_\infty (q/x;q)_\infty (z;q)_\infty
(y/xz;q)_\infty}\;,
\end{array}\end{equation} where the series for
$\phantom{i}_1\Psi_1$ is convergent in
\begin{equation}
\ds
\left|{y\over x}\right|\;<\;|z|\;<\;1\;.
\end{equation}
The results of summations are:
\begin{equation}
\ds\begin{array}{lcl} \ds\mathcal{F}_{y,x}(\N_1,\N_2,\Uh^2) & = &
\ds \g_{1,\lambda}(\N_1,\N_1-\s_y)\,\g_{1,\lambda}(\N_2,\N_2-\s_x)\times\\
&&\\
&&\ds\times \phantom{i}_1\Psi_1(q^{\s_y-\N_1},
q^{1+\N_2};q^{1/2+\N_2-\s_y}{y\over x})\;.\\
\mbox{and}&&\\
&&\\ \ds\widetilde{\mathcal{F}}_{x,y}(\N_1,\N_2,\Uh^2) & = &
\ds \g_{1,\lambda}(\N_1,\N_1-\s_x)\,\g_{1,\lambda}(\N_2,\N_2-\s_y)\times\\
&&\\
&&\ds\times
\phantom{i}_1\Psi_1(q^{\s_y-\N_2},q^{1+\N_1};q^{1/2+\N_1-\s_y}{y\over
x})\;.
\end{array}\end{equation}
Put now $y^2=q^{-1}x^2$, then it appeared
\begin{equation}
\ds \mathcal{F}_{1,2}(\N_1+\N_2,\s_x)\;\stackrel{def}{=}\;
\mathcal{F}_{y,x}(\N_1,N_2,\Uh^2)\;=\;-\;
\widetilde{\mathcal{F}}_{x,y}(\N_1,\N_2,\Uh^2)\;,
\end{equation}
where $\ds y^2=q^{-1}x^2$, and
\begin{equation}\label{Fs}
\ds\begin{array}{l} \ds\mathcal{F}_{1,2}(\N_1+\N_2,\s_x)\;=\;
q^{\N_2(\N_2-\s_x)+\N_1(\N_1-\s_x+1)}\times\\
\\
\ds \times\lambda^{\N_1+\N_2-2\s_x+1} \Theta(q^{\N_2-\N_1})\,
{(q^{2+\N_1+\N_2-\s_x};q)_\infty^{}\over (q;q)_\infty^2}\;.\\
\end{array}\end{equation}
Here it is used the $\theta$-function notation
\begin{equation}
\Theta(x)\;=\;(x;q)_\infty\,(qx^{-1};q)_\infty\,(q;q)_\infty\;=\;
\sum_{n\in\,\mathbb{Z}}\;(-x)^n\;q^{n(n-1)/2}\;,
\end{equation}
such as
\begin{equation}
\ds \Theta(q^k x)\;=\;(-x)^{-k}\,q^{-k(k-1)/2}\;\Theta(x)\;,\;\;\;
\Theta(x^{-1})\;=\;-x^{-1}\,\Theta(x)\;,
\end{equation}
Indeed, due to the equidistance of $\N_1$ and $\N_2$, the
$\mathcal{F}_{1,2}$ depends only on $\N_1+\N_2$.

Now we may calculate the wronskian. By definition, it is
\begin{equation}\label{W12}
\ds
W(x^2)_{1,2}\;=\;Q_1(q^{-1}x^2)\;Q_2(x^2)\;-\;Q_1(x^2)\;Q_2(q^{-1}x^2)\;.
\end{equation}
Considering the monodromies of $Q_1$ and $Q_2$, standing in the
definition of the wronskian and using eq. (\ref{pseudocom}), one
may see that the most parts in the subtraction (\ref{W12}) are
canceled. Only possible exception is the subspace
$\delta_W\;:\;\a_1^+\a_2^{}\;=\;-1$. So to calculate the
wronskian, one has to take a trace only over this subspace. In
general, let $\zeta_1$ and $\zeta_2$ be the characteristics of
$\N_1$ and $\N_2$ respectively. Then using  the definition of
$\mathcal{F}$ and $\widetilde{\mathcal{F}}$, eqs. (\ref{Fs})
equivalence of $\mathcal{F}$ and $\widetilde{\mathcal{F}}$, one
may conclude:
\begin{equation}
\ds
W_{1,2}\;=\;\Xi_{1,2}\;\sum_{n_f\in\zeta_1+\zeta_2+\mathbb{Z}}\;\;
\left(\prod_f\;\mathcal{F}_{1,2}(n_f,\s_f)\right)\;
\left(\prod_f\;(\Uh\Vh)_f^{n_f-n_{f-1}}\right)\;,
\end{equation}
where $\Xi_{1,2}$ is an extra multiplier that may come from the
subtraction, $\delta_W$-trace definition and so on. Nevertheless,
considering the case $\s_x\;=\;\zeta_1$ $modulo$ $\mathbb{Z}$ and
$\zeta_2=0$, one obtains the following useful form of
$\mathcal{F}_{1,2}$:
\begin{equation}
\ds \mathcal{F}_{1,2}(n+\s-1,\s)\;=\;
(-)^n\;q^{n(n-1)/2}\;{\lambda^n\over (q;q)_n}\;\;
\lambda^{-\s}\;{\Theta(q^\s)\over (q;q)_\infty}\;.
\end{equation}
Extracting now the shift operator as it is described in eq.
(\ref{P-trick}), one obtains relation (\ref{W}). Extra multiplier
is equal to unity, this we have checked by a series expansion
with respect to $\lambda$.

\section{Discussion}

The technique and results, given in this paper, are rather formal.
We have dealt with the single Weyl pair in each site of the
lattice, and $q$ is an arbitrary complex number inside the unit
circle. It is well known, this regime is absolutely non-physical,
and thus the results presented are to be considered as just an
exercise in the field of $q$-combinatorial analysis. But,
nevertheless, some applications of the results and technique
presented may be found.

Talking about the Weyl algebra, people usually keep in mind two
aspects: the first one implies the dualization and $\ds q \;=\;
\exp\{i\,\pi\,\EXP^{i\theta}\}$, and the second one is the finite
state $\ds q\;=\;\EXP^{2\,\pi\,i/ N}$. \footnote{This paper
suggests the third aspect, applied in the backward direction yet:
several Toda-chain-type models, physical as well, may be obtained
from a model with arbitrary $q$ in the limit $q\mapsto 1+\hbar$,
regarded in a special way, such that a rational Weyl algebra
mapping is linearized with respect to one of Weyl generators in
the first order of $\hbar$.} Our experience in the Weyl algebra
exercises says that most our results, especially contained the
$q$-dilogarithms and permutations, may be immediately rewritten
in the dualized form. In this way the results may be applied to
the physical relativistic Toda chain, \cite{Leb}. It will be done
in a separate paper.

The second aspect is also valid, especially in the part of the
technique derived. Preliminary considerations show that at the
root of unity the model contains the Baxter curve for the Chiral
Potts model, the point on Baxter's curve is the spectral
parameter of $Q$-operator, our constant parameter $\lambda$ is
connected with the modulus of Baxter's curve. Remarkable is that
in the relativistic Toda chain at the root of unity there appears
only one point at Baxter's curve, while in the Chiral Potts model
such point lives at each site on the spin chain. This fact makes
the relativistic Toda chain much more simple than CPM itself.
This model will be considered in a separate paper.

\noindent {\bf Acknowledgements} Authors would like to thank A.
Isaev, P. Pyatov, S. Pakuliak and A.Volkov for useful discussions.
The work of G.P. was supported in part by ESPIRIT project NTCONGS.

\end{document}